# Multi-modal on-chip nanoscopy and quantitative phase imaging reveals the morphology of liver sinusoidal endothelial cells

**David A. Coucheron[1,‡], Ankit Butola[1,2,‡], Karolina Szafranska[3], Azeem Ahmad[1,2], Jean-Claude Tinguely[1], Peter McCourt[3], Paramasivam Senthilkumaran[2], Dalip Singh Mehta[2] and Balpreet Singh Ahluwalia[1, 4*]**

[1]Department of Physics and Technology, UiT The Arctic University of Norway, Norway.
[2]Bio-photonics and Green Photonics Laboratory, Department of Physics, Indian Institute of Technology Delhi, Hauz-Khas, New Delhi- 110016, India.
[3]Faculty of Health Sciences, Department of Medical Biology, Vascular Biology Research Group, UiT The Arctic University of Norway, 9037 Tromsø, Norway
[4]Department of Clinical Science, Intervention and Technology Karolinska Institutet, Stockholm, Sweden.
‡ These authors contributed equally to this work
*Balpreet.singh.ahluwalia@uit.no

**Abstract:**
Visualization of three-dimensional (3D) morphological changes in the subcellular structures of a biological specimen is one of the greatest challenges in life science. Despite conspicuous refinements in optical nanoscopy, determination of quantitative changes in subcellular structure, i.e., size and thickness, remains elusive. We present an integrated chip-based optical nanoscopy set-up that provides a lateral optical resolution of 61 nm combined with a highly sensitive quantitative phase microscopy (QPM) system with a spatial phase sensitivity of ±20 mrad. We use the system to obtain the 3D morphology of liver sinusoidal endothelial cells (LSECs) combined with super-resolved spatial information. LSECs have a unique morphology with small nanopores (30-200 nm in diameter) that are present in the plasma membrane, called fenestration. The fenestrations are grouped in discrete clusters called sieve plates, which are around 100 nm thick. Thus, imaging and quantification of fenestration and sieve plate thickness requires resolution and sensitivity of sub-100 nm along both lateral and axial directions. In the chip-based nanoscope, the optical waveguides are used both for hosting and illuminating the sample. A strong evanescent field is generated on top of the waveguide surface for single molecule fluorescence excitation. The fluorescence signal is captured by an upright microscope, which is converted into a Linnik-type interferometer to sequentially acquire both super-resolved images and quantitative phase information of the sample. The multi-modal microscope, when operated in nanoscopy mode, provided an estimate of the fenestration diameter of 124±41 nm and in QPM mode revealed the average estimated thickness of the sieve plates in the range of 91.2±43.5 nm for two different cells. The combination of these techniques offers visualization of both the lateral size (using nanoscopy) and the thickness map of sieve plates, i.e. discrete clusters fenestrations in QPM mode.

**Introduction**

Far-field optical nanoscopy techniques are frequently used to visualize subcellular structures in biological specimens by surpassing the diffraction limit. Optical nanoscopy encompasses a plethora of techniques, including stimulated emission depletion (STED) microscopy[1]; structured illumination microscopy (SIM)[2]; different variants of single molecule localization microscopy (SMLM) such as photo-activated localization microscopy (PALM)[3] and direct stochastic optical reconstruction microscopy (dSTORM)[4]; and intensity fluctuation based techniques such as super-resolution optical fluctuation imaging (SOFI)[5]. These techniques can help detect subcellular structures (<200 nm) of biological specimens such as lipids, proteins, membrane structures, microtubules and nucleic acids by specific fluorescence tagging[6]. Each technique has respective advantages and disadvantages. SIM has gained popularity for live-cell imaging due to its fast image acquisition time, but at a limited resolution[7]. *d*STORM, on the other hand, is slower, but offers extremely high resolution for e.g. characterization of viral proteins[8] and imaging actin filaments in mammalian cells[9,10]. A lot of effort is spent on developing new super-resolution techniques for a wide range of applications.

To reduce the complexity of the typical SMLM setup using total internal reflection fluorescence (TIRF) configuration, a photonic chip-based optical nanoscopy system was recently proposed[11–13]. The main advantage of the chip-based system is the decoupling of the excitation and collection pathways, as well as miniaturization of the excitation light path of the system. In chip-based nanoscopy, the TIRF illumination is generated through the evanescent field of waveguides, rather than using conventional high magnification and high numerical aperture (N.A.) TIRF lens. The evanescent field in waveguides can be generated over extraordinarily large areas, as it is only defined by the waveguide geometry. This makes it possible to use any imaging objective lens to image arbitrarily large areas as compared to the traditional TIRF-based *d*STORM[12], which is limited by the field of view of the TIRF lens.

Although all these optical nanoscopy techniques offer nanometric spatial resolution they cannot provide the complete morphology, i.e. quantitative change (change in refractive index and thickness) in the subcellular structure. Additionally, a complete 3D image of a cell is required to understand the complete quantitative information. Quantitative phase microscopy (QPM) is an emerging label-free optical technique which facilitates highly sensitive measurements of the refractive index and thickness of both industrial and biological specimens[14]. Various QPM methods have been proposed so far for extracting optical phase and dynamics of biological cells[15–20]. These techniques offer high spatial and temporal phase sensitivity, transverse resolution and acquisition rate[15]. The spatial and temporal phase sensitivity of the QPM system is highly dependent on the illumination source and the type of interferometric geometry, respectively[14,21]. For example, common path QPM techniques offer better stability and temporal phase sensitivity which can be used to measure membrane fluctuation of the cells[22]. Additionally, the spatial phase sensitivity of the system can be improved by using low coherence light sources (halogen lamps and LED), but this requires phase shifting techniques to utilize the whole field of view of the camera[23]. A recent advancement in the QPM technique is used to offer three-dimensional information of the samples by measuring the phase across multiple angles of illumination[17]. This technique offers tomography of various biological specimens such as red blood cells (RBC), HT29 cells and bovine embryos[17,19]. Since the lateral resolution of these techniques depends on the numerical aperture (NA) of the objective lens, imaging beyond this limit (<200 nm) is still challenging and limits its application to the study of subcellular structures. Therefore, it is useful to develop multi-modality routes where different microscopy methods can be used to provide complementary information about the specimen.

For several biological samples, e.g. liver sinusoidal endothelial cells (LSECs), it is useful to extract both the functional information from fluorescent tagging and the quantitative 3D morphological information. LSECs contain large numbers of fenestrations (transcellular nanopores) in the plasma membrane, typically clustered in groups of 10-50 within areas called sieve plates[24]. The fenestrations act as an ultra-filter between blood and the underlying hepatocytes, facilitating the bi-directional exchange of substrates between hepatocytes and blood. For example, smaller viruses and drugs can pass this barrier while blood cells are retained within the sinusoidal vessel lumen[25,26]. Figure 1 presents a schematic of a typical LSEC. The diameter of the fenestrations (white spots in Fig. 1(a)) varies from 50-200 nm[24,26–28], smaller than the diffraction limit of optical microscopy. These fenestrations are often clustered in sieve plates and connected by actin fibers to other sieve plates as shown in Fig. 1(a). The typical thickness of sieve plates is around 100-150 nm[29], and these fenestrations are consequently nanoscale-sized in all the three dimensions. As shown in Figure 1, the fenestrations and the sieve plates are located in the LSEC's plasma membrane and therefore TIRF illumination is ideally suited for imaging these structures. Determining the diameter and height of fenestrated regions can be important as it can be affected by several drugs[30–32], as well as during natural (but detrimental) changes such as aging that result in "pseudocapillarisation", whereby LSEC simultaneously lose fenestrations and become thicker[33].

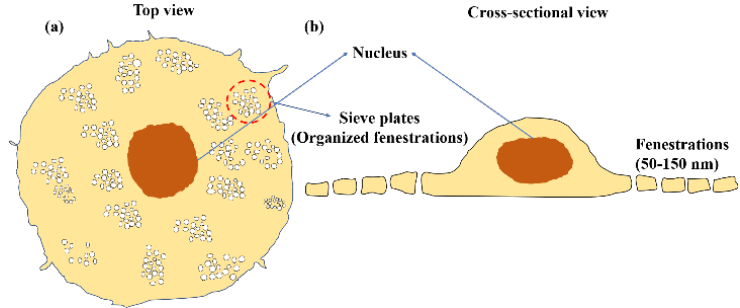

Figure 1: Top view (a) and cross-sectional view (b) of liver sinusoidal endothelial cells. LSECs have unique morphology, where nanoscopic fenestrations are grouped in thin sieve plates. The diameter and the thickness of fenestrations are below the diffraction limit of conventional optical microscopes.

Here, we have developed a multi-modal chip-based optical nanscopy and highly sensitive QPM system to visualize three-dimensional morphological changes in LSECs. Although, QPM offers nanometric sensitivity in the axial direction, it cannot exactly locate these fenestrations due to its diffraction limited lateral resolution. Interestingly, the nanometric sensitivity of QPM technique in axial direction can be utilized to find the optical thickness of sieve plates if one can locate the fenestrations very precisely. On the other hand, chip-based $d$STORM supports super-resolution imaging down to 50 nm over an extraordinarily large field of view (FOV)[11]. The proposed system enables decoupling the light illumination path from the collection path and thus relatively easy integration of $d$STORM and QPM. Moreover, it allows super resolution imaging in the lateral dimension (with $d$STORM) and nanometric sensitivity in the axial direction (with QPM) giving the complete 3D morphology of the cells. Finally, the system offers a combination of simultaneous functional and quantitative imaging. In this work we demonstrate the capabilities of the system by imaging LSECs with both diffraction limited TIRF microscopy and $d$STORM. The plasma membrane fenestrations and sieve plates are observable with $d$STORM and the average optical thickness of the sieve plate region is obtained using QPM. The schematic diagram of the integrated chip-based nanoscopy and QPM setup is shown in Fig. 2. A Cobolt Flamenco laser at 660 nm is coupled into the waveguide to generate the evanescent field on top of the waveguide to perform the $d$STORM experiment. QPM is performed by using 560 nm Cobolt laser. Further details are provided in the Methods section.

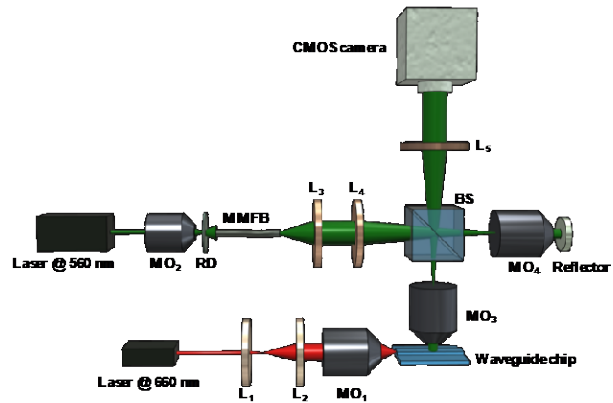

Figure 2: Schematic diagram of integrated partially spatially incoherent quantitative phase microscopy (QPM) and chip-based nanoscopy system for the morphological imaging of liver sinusoidal endothelial cells. $MO_{1-4}$: Microscopic objective lens, RD: Rotating diffuser, $L_{1-5}$: Lens, MMFB: Multi-multimode fiber bundle, BS: Beam splitter. High intensity evanescent field is generated on top of the waveguide chip using 660 nm Cobolt laser for single molecule fluorescence excitation. The fluorescence signal is captured by an upright microscope which is converted into a Linnik type interferometer to perform quantitative phase microscopy (QPM).

**Results and Discussion**

The spatial phase sensitivity represents the spatial noise present in the QPM system, which plays an important role in determining an accurate phase map of thin samples. To measure the phase noise in the system, we imaged a standard flat mirror of surface flatness $\lambda/10$. Figure 3(a) shows the recorded interferogram on the mirror surface when operating the system in QPM mode. Ideally, the measured phase should be zero, but it is not due to spatial phase noise present in the system. Figure 3(b) represents the standard deviation of the phase variations, i.e., the spatial noise of the system. The average spatial noise of the system was ±20 mrad, which is significantly less than using a direct laser to perform QPM[34]. High spatial and temporal coherence of a direct laser causes speckles and spurious fringes in the final image, reducing the phase sensitivity of the QPM system. This unwanted noise can be avoided by introducing spatial and temporal diversity in the laser beam by passing it through a rotating diffuser and subsequently a multi-multimode fiber bundle (MMFB)[35]. The rotating diffuser and MMFB reduce the spatial coherence of the light source, thus improving the spatial phase sensitivity of the system.

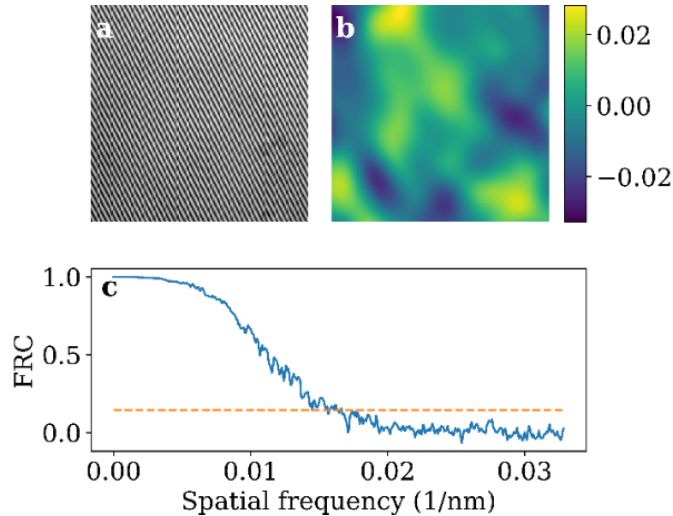

Figure 3: Noise characterization of the QPM system. (a) Interferogram captured by the QPM mode of the proposed setup on a standard mirror of $\lambda/10$ surface flatness. (b) Standard deviation of phase in (a) demonstrating the phase noise of the system. The color bar represents the phase map in radians. (c) The 61 nm lateral optical resolution of the chip-based system was obtained on the sample used in Figure 4 using Fourier ring correlation.

Figure 4 shows a complete dataset gathered for one imaged region of LSECs. It consists of: a) a bright field image, b) a phase map of the LSECs, c) a diffraction limited TIRF image and d) a *d*STORM image with visible fenestrations. The figure shows parts of three different cells. The higher phase region represents the nucleus, with the plasma membrane surrounding it. The maximum phase value is 2.3 rad in the nucleus of the bottom left cell. The total dataset gives nanometric sensitivity in the axial direction, together with super-resolution in the lateral direction from the *d*STORM. Fig. 4 e)-g) present the inset from d) in TIRF, *d*STORM and QPM mode. The phase of the imaged membrane region varies from 0 to approximately 0.25 rad. Comparing Figure 3e and 3f, the fenestrations in the membrane are clearly resolved in the *d*STORM image, not being visible using diffraction-limited TIRF imaging.

A Fourier ring correlation test was performed on the *d*STORM data from Figure 4 to estimate the resolution, with the result correlation plotted in Figure 3b. The resolution at FRC=1/7 is 61 nm. The resolution of our system can, however, be increased by swapping the beam splitter (marked BS in Figure 2) in the system for a flip mirror, as half of the photons are lost passing through it. We chose to use a beam splitter as it opens up for simultaneous fluorescence and phase imaging, whereas with a flip mirror it would be limited to sequential imaging.

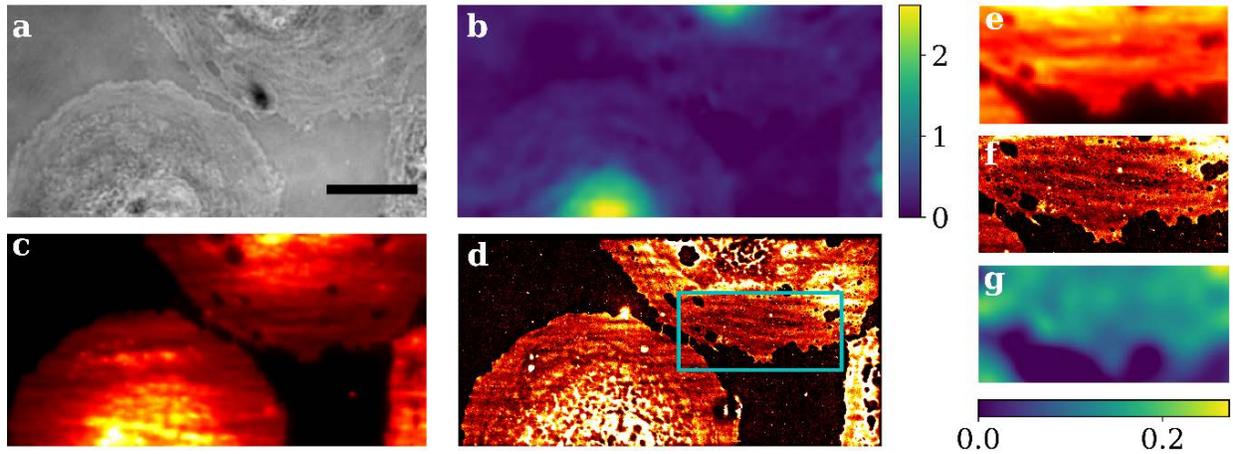

Figure 4: Parts of three cells imaged with brightfield (scalebar is 10 μm) (a), QPM (b), TIRF (c) and dSTORM (d). The phase map gives morphological information about the cells, with a maximum phase value in the nucleus of the lower left cell of 2.3 rad. The dSTORM image clearly shows plasma membrane fenestrations in the top cell. TIRF, dSTORM and QPM images of the inset in (d) is also presented in (e-g). The colour bars show phase in radians.

LSECs have an interesting morphology with a thick nucleus in the center and an extremely flat cell membrane containing nanoscale fenestrations grouped within sieve plates. We used dSTORM to locate fenestrated regions in LSECs and QPM to determine the average phase of the fenestrated regions. Figure 5 shows a dSTORM image and the corresponding phase image. Figure 6 shows the measured phase value for several different fenestrated regions for two different cells in Figure 5. Although the fenestrations are below the diffraction limit, and thus the spatial resolution limit of the QPM, the average phase of the sieve plates can be calculated. The average optical thickness of the sieve plates can then be calculated based on the phase map. The calculation of the actual thickness does however require an estimate for the effective refractive index. Sample 1 from Figure 5 has a weighted average phase over all regions of 0.158 rad with a standard deviation of 0.041 rad. Sample 2 has a weighted average phase of 0.131 rad with a standard deviation of 0.045 rad. The values are thus in agreement. If we assume a refractive index range of 1.35 – 1.37[36] using equation 4 we get an average thickness of 92.5±39.1 nm and 76.3±36.4 nm for sample 1 and 2, respectively. Combining the data, the resulting height is 91.2±43.5 nm. We also calculated the diameter of the fenestrations of the sample in Figure 5a using a local thresholding algorithm. The mean diameter of the fenestration is 124 nm with a standard deviation of 41 nm. Both the thickness and the pore diameter agrees well with previous literature[29].

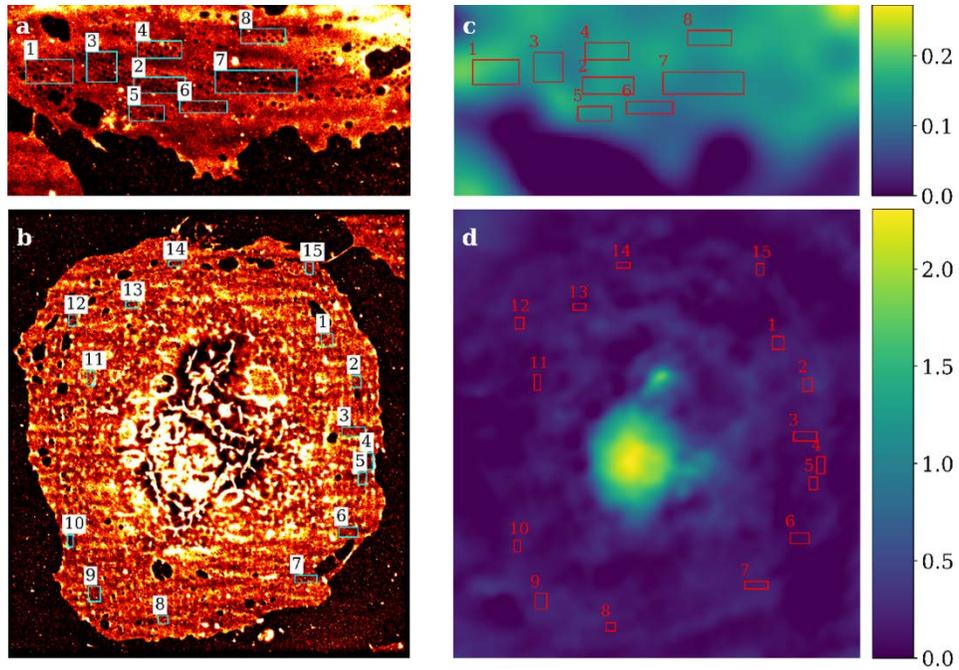

Figure 5: (a,c) *d*STORM and phase image of the inset in Figure 3. (b,d) *d*STORM and phase image of an entire cell. Fenestration in the plasma membrane is visible all throughout the cell. The phase shows a maximum phase of 2.3 rad in the nucleus of the cell. The color bars show phase in radians.

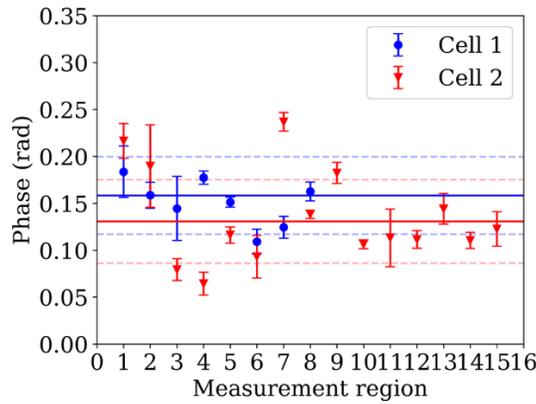

Figure 6 Phase value measured for the marked regions in Fig. 5 (a) and (b) with standard deviation of each region. The mean values for all regions are presented as a line for both cells, with the standard deviation of the regions presented as dotted lines.

**Conclusion**

In this work we have developed a muti-modal chip-based optical super-resolution microscopy and highly spatially sensitive quantitative phase microscopy system. To demonstrate the system's potential we localized plasma membrane fenestrations in liver sinusoidal endothelial cells using *d*STORM and then measured the thickness of the fenestrated areas using QPM. The system, when operated in the *d*STORM mode, offers nanometric spatial resolution (61 nm) to visualize small fenestrations present in LSECs. Further, precise localization of the group of fenestrations and the improvement in phase sensitivity offers the optical thickness of sieve plates. Assuming an average refractive index of the cell membrane, the measured average thickness of the fenestrated regions were 92.5±39.1 nm and 76.3±36.4 nm for sample 1 and 2, respectively. Furthermore, the diameter of the fenestration was found to be 124±41 nm.

The system enables multi-modal imaging in a simple manner, while still being easy to further customize. For improved resolution in *d*STORM, a simple flip mirror instead of a beam splitter will help[11]. Further enhancement in phase sensitivity is possible by replacing the partial spatial coherent illumination with a perfectly incoherent light source such as white light or a light emitting diode (LED). The white light source offers maximum possible spatial phase sensitivity[37], but requires multiple frames, i.e. phase shifting interferometry (PSI) to extract the phase information due to poor temporal coherence. Additionally, PSI can also be useful to improve the transverse resolution of the system. Moreover, with minor modifications, different modalities can easily be added with the current system, such as waveguide based optical trapping[38] and spectroscopic techniques[39,40].

Chip-based microscopy has been implemented for live cell imaging of delicate cells[41,42]. In future work we aim to adapt the proposed multi-modality microscopy platform for imaging dynamics of fenestration in living LSECs e.g. when challenged by chemicals or drugs that alter the fenestrations and sieve plates. Being able to get both fenestration diameter and sieve plate thickness makes it possible to track changes in a very detailed manner. This will be a particularly useful tool for the discovery of agents that reverse age-related pseudocapillarisation[33] since the method simultaneously measures two important parameters, LSEC thickness and fenestration, that are increased and reduced (respectively) during the ageing process.

**Methods**

**Experimental set-up**

**Working principle of quantitative phase microscope:** The schematic diagram of partially spatially coherent QPM setup is shown in Fig. 2. A highly coherent Cobolt laser (@ 560 nm) light is expanded by a microscopic objective ($MO_2$) and passes through the rotating diffuser and MMFB. The output beam of the MMFB acts as a partially spatially coherent and temporally coherent source due to spatial and temporal diversity [a]. It has been shown previously that the reduction of spatial coherence results in speckle free images and improves the spatial phase sensitivity of the interferometry system[20,35]. Therefore, partially spatially coherent sources can be utilized to extract the morphological changes of thinnest biological specimens such as LSECs. The partially spatially coherent beam further coupled into the Linnik type QPM system. In the QPM system, light beams reflected from the sample and reference mirror interfere at the beam splitter plane. The 2D interference pattern coded the information of the sample which further captured by the CMOS image sensor (Hamamatsu ORCA-Flash4.0 LT, C11440-42U).

The 2D intensity distribution of the interferogram can be expressed as:

$$I(x,y) = a(x,y) + b(x,y)cos[2\pi i(f_x x + f_y y) + \phi(x,y)] \qquad (1)$$

where a(x,y) and b(x,y) represent the background and the modulation terms, respectively. $f_x x$ and $f_y y$ are the spatial frequencies of the interference pattern along x and y directions and $\phi$ (x,y) is the phase difference between the object and reference beam.

Standard Fourier transform analysis[43] and Goldstein phase unwrapping algorithm[44] are used to extract the phase information of the specimens. The phase information is a combination of refractive index and thickness of the specimens and can be written as:

$$\phi(x,y) = \frac{2\pi}{\lambda} \times 2h(x,y) * \{n_s(x,y) - n_0(x,y)\} \qquad (2)$$

where λ is the wavelength of incident light, h is the geometrical thickness of the specimen; $n_s$ and $n_o$ are the refractive indices of the specimen and surrounding medium, respectively and an extra factor of 2 appears because the imaging is performed in the reflection mode. By reformatting the equation an expression for the thickness of the sample can be found:

$$h(x,y) = \frac{\lambda * \phi(x,y)}{4\pi * \{n_s(x,y) - n_0(x,y)\}} \tag{3}$$

**Chip preparation**: All imaging in this work was done using $Si_3N_4$ strip waveguides with varying widths between 200 and 500 μm. The chips were fabricated using a previously described procedure[42]. Before any sample preparation, the chips were thoroughly cleaned using a two-step process. The chips were first cleaned in a 1% Hellmanex in deionized (DI) water at 70 °C for 10 minutes. Following that, the chips were rinsed with DI water, followed by isopropanol and DI water again. Finally, the chips were dried using $N_2$. A hollow rectangular chamber was created with polydimethylsiloxane (PDMS) and placed on the chip to restrict the area where cells attach.

**Cell isolation and seeding**: Cell have been isolated from C57BL/6 male mouse using modified standard protocol[45]. Briefly, perfusion of the liver with Liberase™ (Roche) was followed by low speed differential centrifugation and then separated using superferromagnetic beads conjugated with the LSEC-specific antibody CD146 (MACS, Miltenyi Biotec). After isolation the cells were seeded on chips precoated with human fibronectin and incubated in 5% $CO_2$ at 37°C in RPMI-1640 culture medium for 2 hours. Seeding density was about 100 000 LSECs per $0.5cm^2$ PDMS chamber. Samples were fixed by 10 minutes incubation in 4% PFA in PBS and left in 1% PFA at 4°C until imaging.

**Staining protocol:** The cells were all stained with CellMask Deep Red (CMDR). The chips were rinsed thoroughly with PBS before staining. A 1:1000 dilution of CMDR in PBS water was added inside the PDMS chamber and left to incubate for 10 minutes. The sample was then thoroughly rinsed with PBS again. Prior to imaging, a dSTORM buffer was prepared using 22.5 μL PBS, 22.5 μL $H_2O$-based oxygen scavenger system solution[46], and 5 μL 1M MEA. The sample was then rinsed thoroughly with PBS before the blinking buffer was applied and the sample area sealed off with a cover slip.

***d*STORM imaging and data analysis:** *d*STORM imaging was done using chip-based TIRF excitation. Once the sample was stained and the blinking buffer added, the chip was placed on the sample stage and held in place with a vacuum chuck. The excitation light was coupled from free space by end-fire coupling using a 0.5 NA objective lens. The waveguides are multi-moded, giving rise to an inhomogeneous excitation pattern. In order to achieve homogeneous resolution, the coupling objective was scanned along the input facet to average out the modes. Imaging was done with a Hamamatsu Orca sCMOS camera with 30 ms exposure time. For TIRF images, the exposure time was increased to 100 ms and an average of approximately 1000 frames used. Approximately 100 mW power was used for all images, however the power was incrementally increased up to approximately 400 mW towards the end of each imaging procedure to get additional localizations. The data were reconstructed using ThunderSTORM[47], a FIJI plugin,

**Author Contributions**

D.A.C. and A.B. contributed equally to this work. D A.C, A.B. and B.S.A. wrote the paper and all other authors contributed in the paper revision. All authors have given approval to the final version of the manuscript. D A.C. did the dSTORM experiments and its analyses, A. B. did the QPM experiments and its analyses, J.C.T. designed the waveguides, D.A.C, A.B. A.A. and J.C.T. developed the experimental set-up. K.S. isolated the LSECs. K.S. and. P.McC. assisted with biology. P.S. and D.S.M. assisted with the QPM. B.S.A. conceived the idea, supervised the project and provided funding resources to this project.

**Funding Sources**

B.S.A. acknowledges UiT, The Arctic University of Norway Tematiske Satsinger funding program and Diku - Direktoratet for internasjonalisering og kvalitetsutvikling i høyere utdanning (Project number INCP-2014/10024) and the European Union's Horizon 2020 research and innovation program under the Marie Sklodowska-Curie Grant Agreement No. 766181, project "DeLIVER".

**ACKNOWLEDGMENT**